# Kinetics of solubilization with Triton X-100 of egg-yolk lecithin bilayers containing cholesterol

ŞTEFAN HOBAI, ZITA FAZAKAS

*Department of Biochemistry, University of Medicine and Pharmacy, Tg. Mureş, Romania*

***Correspondence address:***
*Ştefan Hobai, Department of Biochemistry, University of Medicine and Pharmacy, Târgu Mureş, Romania. Fax: 040-065-210407, E-mail: fazy@netsoft.ro*

**Abstract**

The titration solubilization of multilamellar egg-yolk lecithin liposomes (MLV-EYL) with Triton X-100 was studied by rectangular optical diffusimetric measurements as a function of cholesterol (Chol) concentration. It was determinated the variation of optic percentage diffusion (per μmol surfactant), $\Delta Dif\%/\mu mol\ _{TX-100}$, in the course of solubilization of MLV-EYL-Chol system with TX-100 10mM. The statistical analysis of the titration curves can reveal the contribution of cholesterol to the stability of phospholipid bilayer membranes. The solubilization of the lecithin-cholesterol mixtures, with a high cholesterol content, much more bile salt requires.

**Introduction**

The interaction of surfactants with phospholipid liposomes produces the disintegration of bilayers and formation different non-lamellar structures as micelles. Information on kinetics of liposome solubilization with surfactants is scarce. However there are some papers concerning the kinetics of interaction of phosphatidylcholine liposomes with Triton X-100 [Alonso,1987] with sodium deoxycholate [Lichtenberg,1979] or inversely, the kinetics of *micellar → lamellar* phase transition in phosphatidylcholine-cholate mixtures [Lichtenberg, 1989]. There are some researches on the carboxyfluorescein or calcin studies release by solubilization with surfactants as Triton X-100 and sodium dodecylsulphate [Maza,1996; Gregoriadis,1984]. The effect of cholesterol on kinetics of multilamellar vesicle solubilization with Triton X-100 may be used to explain some biological fenomena as the major role played by it in the bile and gallstone [Müller, 1990;Bondar,1998]. The statistical analysis of the titration curves, obtained by rectangular diffusimetric measurements, can reveal the contribution of cholesterol to the stability of phospholipid bilayer membranes.

**Apparatus, materials and methods**

Egg yolk lecithine (EYL) was purchased from Sigma Chemical Co. type X-E (cat.no.P5394/1996) purified by us by neutral alumina column chromatography, verified by thin layer chromatography (silicagel). The mobile phase was a mixture $CHCl_3:CH_3OH:H_2O$ (65:30:4,vol). The identification of the phospholipid was made with iodine and shows a single spot. 1 ml EYL stock solution of 18.6mM concentration in a solvent mixture $CHCl_3:CH_3OH$ (9:1,vol) was evaporated under methane stream for 15 minutes, at 52 °C. The lipid film formed on the wall of a 100 ml flask attached to the rotary evaporator, was dried at a low pressure (p<0.01 Torr) for 4 hours at room temperature. Taking into account the mass of the lipid film and phosphate content of it (mineralized with nitroperchloric mixture,and assayed by Briggs method), we determined the mean molecular weight of the lecithin, as being 766 Da. This molecular weight is almost identical with the value of the lecithine supplied by Avanti Polar Lipids Inc.

The Triton X-100 (TX-100) was purchased from Merck, from Sigma Chemical Co. and Cholesterol (Chol) from Fluka AG. Tris(hydroxymethyl)-aminomethane- HCl (Tris-Cl) was purchased from Austranal. These chemicals and the organic solvents were of analytical grade. Double distilled water was used.

By using a spectrophotometer Specol with FR optical diffusion system and photocells and voltage source type HQE 40 and UV lamp type OSRAM there were titrated at a 0,04µmol/s rate at 20°C, samples of 1.4ml suspension (0.7µmols of EYL) which vesicles contained cholesterol at different molar ratios with phospholipid.

**Liposome dispersion preparation**

Appropriate amounts of EYL in chloroform:methanol (9:1,vol) and cholesterol in chloroform solutions were evaporated, in a glass test tube, under a nitrogen stream until it reached the consistence of a lipid film on the wall tube settled down. The film was purged with the gas for 30 minutes in order to eliminate the alcohol traces. The film hydration and bilayer suspension were made with buffer Tris-Cl 0.05M, pH=7.2 at 0.6mg/ml (lipid/solution) by vortexing 10 s.

**Results and Discussion**

The rectangular optical diffusion data are presented in Table 1. The formula used to estimate the statistical significance of these data are based on the Student test (t). $\overline{X}_1$ and $\overline{X}_2$ are the average values of individual values. $n_1$ and $n_2$ are the volumes of two selections and $S_1$ and $S_2$ are standard deviations of two selections.

$$\overline{S}_{12} = \sqrt{\frac{(n_1-1) S_1^2 + (n_2-1)S_2^2}{n_1+n_2-2}} \quad (1)$$

$$t_{12} = \frac{|\overline{X}_1 - \overline{X}_2|}{\overline{S}_{12}\sqrt{1/n_1+1/n_2}} \quad (2)$$

**Table 1.** *The variation of optic percentage diffusion (per µmol added surfactant), ΔDif%/ µmol TX-100, in the course of solubilization of MLV-EYL-Chol system with TX-100 10mM.*

| | ΔDif%/µmol TX-100 | | | | | |
|---|---|---|---|---|---|---|
| | EYL:Chol 1:0 | EYL:Chol 10:1 | EYL:Chol 5:1 | EYL:Chol 3.3:1 | EYL:Chol 2.5:1 | EYL:Chol 1:1 |
| 1 | 34.42 | 14 | 12.25 | 12.25 | 10.55 | 26.66 |
| 2 | 22.94 | 56 | 32.66 | 49 | 47.5 | 26.66 |
| 3 | 34.42 | 22.36 | 32.66 | 24.5 | 31.66 | 21.33 |
| 4 | 68.83 | 37.27 | 32.66 | 24.5 | 23.75 | 21.33 |
| 5 | 45.88 | 28 | 32.66 | 24.5 | 23.75 | 9.7 |
| 6 | 27.53 | 28 | 24.5 | 32.6 | 23.75 | 9.7 |
| 7 | 27.53 | 28 | 16.33 | 19.6 | 19 | 6 |
| 8 | 34.42 | 28 | 24.5 | 19.6 | 13.57 | 4.85 |
| 9 | 19.66 | 28 | 12.25 | 14 | 15.83 | - |
| 10 | 13.76 | 9.32 | 7 | 4.66 | 4.75 | - |

The statistical processing of data in Table 1 is to be seen in Table 2.

**Table 2.** *Statistical parameters of rectangular optical diffusion data of MLV-EYL-Chol system solubilization with TX-100.*

| Nr. | EYL:Chol (molar ratios) | The average of ΔDif%/μmol TX-100 | Standard deviation,S | Volume of selection, n |
|---|---|---|---|---|
| 1 | 1:0 | 37 | 14,6 | 8 |
| 2 | 10:1 | 30,2 | 12,3 | 8 |
| 3 | 5:1 | 26,03 | 8,14 | 8 |
| 4 | 3,3:1 | 25,82 | 11,02 | 8 |
| 5 | 2,5:1 | 24,19 | 11,5 | 8 |
| 6 | 1:1 | 15,78 | 9,16 | 8 |

Table 3 shows the statistical parameters resulting from comparing the titration curves with TX-100 of MLV-EYL suspension with the titration curves (5) of MLV-egg-yolk-cholesterol systems.

**Table 3.** *Statistical parameters of comparison of the titration curves of MLV-EYL (1) with those of MLV-EYL-Chol, having the molar ratios EYL:Chol (2)=10:1; (3)=5:1; (4)=3.3:1; (5)=2.5:1; (6)=1:1.*

| $\overline{S}_{12}$ = 13.5 | $t_{12}$ = 1 | $\nu$=16 | Statistical non-significance | 0.2<P<0.3 |
|---|---|---|---|---|
| $\overline{S}_{13}$ = 11.82 | $t_{13}$ = 1.86 | $\nu$=16 | Statistical significance | 0.1<P<0.2 |
| $\overline{S}_{14}$ = 12.93 | $t_{14}$ = 1.73 | $\nu$=16 | Statistical significance | 0.1<P<0.2 |
| $\overline{S}_{15}$ = 13.14 | $t_{15}$ = 1.95 | $\nu$=16 | Statistical significance | 0.1<P<0.2 |
| $\overline{S}_{16}$ = 13.11 | $t_{16}$ = 2.76 | $\nu$=16 | Good statistical significance | 0.01<P<0.05 |

This comparison reveals the fact that the increase of molar concentration of cholesterol in the phospholipid bilayers slows down their solubilization with surfactants. The best statistical significance can be observed by comparing the MLV-EYL solubilization with the MLV-EYL-cholesterol, at the molar ratio EYL:Chol=1:1. This phenomenon is due to the inclusion of cholesterol in the lipid bilayers, and it modifies the bile salt partition between the bilayers and water environment in favour of the latter as a consequence, in order to obtain a certain effective molar ratio surfactant/phospholipid, $R_e$, it is necessary to have a biliare salt concentration higher when the bilayers contain cholesterol, than in the case when the cholesterol is missing. On the other hand, the solubilization of the lecithin-cholesterol mixtures, with a high cholesterol content , in order to form stable mixed micelles cholesterol-EYL-DOCNa, it requires much more bile salt.

**Glossary**

Chol- cholesterol
Dif%- optic percentage diffusion (per μmol added surfactant)
EYL - egg-yolk lecithin
MLV- multilamellar vesicles
TX-100, Triton X-100, polyoxyethylene-p-t-octilfenol

**References**
1. Alonso,A.,Urbaneja,M.A.,Goni,F.M.,Carmona,F.G.,Canovas,F.G.,Fernandey,C.G.,*Kinetic studies on the interaction of phosphatidylcholine liposomes with Triton X-100*, Biochim.Biophys. Acta, 1987,**902**,237-246
2. Bondar,O.P.,Rowe,E.S.,*Role of cholesterol in the modulation of interdigitation in phosphatidylethanols*, Biochim.Biophys.Acta,1998,**1370**,207-217
3. Gregoriadis,G.(ed.), Liposome Technology,Vol.I,II,III,CRC Press,Boca Raton,FL.,1984
4. Lichtenberg,D., Zilberman,Y.,Greenzaid,P.,Zamir,S.,*Structural and kinetic studies on the solubilization of lecithin by sodium deoxycholate*, Biochemistry,1979,**18**,3517-3525
5. Lichtenberg,D.,Almog,S.,Kushnir,T.,Nir.S.,Structural and Kinetic Aspects of the Micellar ⇔Lamellar Phase Transformations in Phosphatidylcholine-cholate Mixtures, in: Surfactants in Solution,(K.,L.,Mittal,ed), Plenum Publishing Corporation,1989
6. de la Maza,A.,Parra,J.L.,*Changes in phosphatidylcholine liposomes caused by a mixture of Triton X-100 and sodium dodecyl sulfate*, Biochim.Biophys. Acta,1996,**1300**,125-134
7. Müller,K.,Schuster,A.,*Solubilization of multilamellar liposomes of egg yolk lecithin by the bile salt sodiumtaurodeoxycholate and the effect of cholesterol- a rapid-ultrafiltration study*, Chem.Phys.Lipids,1990,**52**,111-127